\documentclass[12pt,letterpaper]{article}
\usepackage{amsfonts,amsmath}
\usepackage{lscape,graphicx,epsf}

\usepackage[headings]{}

\def\ds{\displaystyle}
\def\bea{\begin{eqnarray}}
\def\eea{\end{eqnarray}}
\def\be{\begin{equation}\bea\ds}
\def\ee{\ea\end{equation}}
\def\bmlt{\begin{multline}}
\def\emt{\end{multline}}

\def\gaugeA{{\rm A}_{gauge}}

\def\A{\mathcal{A}}
\def\B{\mathcal{B}}
\def\Z{Z}
\def\N{\mathcal{N}}
\def\F{\mathcal{F}}
\def\M{\mathcal{M}}
\def\we{\wedge}

\def\a{{\textsl a}}

\begin{document}

\begin{titlepage}

\begin{flushright}
{SITP-09/39}\\
{ITEP-TH-28/09}\\
\end{flushright}

\vskip1cm
\begin{center}
{\Huge\bf Flavor brane on the baryonic branch of moduli space} \\
\vskip0.5cm {\Large A.~Dymarsky\footnote{Current address:
School of Natural Sciences, Institute for Advanced Study, Princeton, NJ, 08540}}
\vskip1cm
\large\itshape Stanford Institute for Theoretical Physics,\\
Stanford University, Stanford, CA 94305, USA\\
\end{center}

\vspace{0.5cm}
\begin{abstract}
We study an extra flavor in the cascading $SU((k+1)M)\times SU(k M)$ gauge theory by adding probe D7-brane to the geometry. By finding a solution to the kappa-symmetry equation we establish that the D7-brane is mutually supersymmetric with the background everywhere on the baryonic branch of moduli space. We also discuss possible applications of this result.
\end{abstract}

\end{titlepage}

\section{Introduction}
\label{intro}

The cascading $SU((k+1)M)\times SU(k M)$ theory of \cite{KS} provides an interesting example of supersymmetric field theory with a rich structure of moduli space. The low energy superpotential constraints the colorless baryon $\A,\B$ and meson $\M$ fields to satisfy
\bea
\label{branchconstraint}
\A\B-{\rm det}\M=\Lambda_{2M}^{4M}\ .
\eea
The baryonic branch of moduli space follows from (\ref{branchconstraint}) with $\M=0$ and generalizes the $\Z_2$ invariant vacuum  $\A=\B$ dual to the famous Klebanov-Strassler solution \cite{KS}. The latter can be generalized to the one-dimensional family of supergravity backgrounds \cite{GHK} that were found in \cite{BGMPZ}. Their  corresponding geometries are sometimes called resolved warped deformed conifolds because they break the $\Z_2$ symmetry of the KS solution which exchanges the two $S^2$ of the deformed conifold.  These solutions are dual to the field theory on the baryonic branch of moduli space and could be called the baryonic branch of the KS solution.

The field theory in question admits addition of fundamental matter without breaking the supersymmetry of the Lagrangian.
Potentially, the full moduli space of the original theory (\ref{branchconstraint}) may not survive hence posing a question of finding the moduli space of the new theory. Besides obvious field theory motivation this is an interesting question in the context of the warped throat compactifications. We refer the reader to the Discussion section for details while in the rest of the paper we deal with this question in one particular example of the field theory introduced in \cite{Kuperstein}.

In principle the question of finding the moduli space can be addressed by analyzing the field theory superpotential.
The fundamental matter adds an extra term to the superpotential of the original theory and usually this extra term vanishes if the fundamental fields do. For example in the field theory in question the extra term in the superpotential is \cite{Ouyang,Kuperstein}
\bea
\sim q(A_1B_1+A_2B_2)\tilde q\ .
\eea
Clearly in the perturbative regime there is always a supersymmetric solution $q=\tilde q= 0$ for any value of baryon fields $\A,\B$. Consequently one is to conclude that the whole baryonic branch of the moduli space survives after fundamental matter is added. We note though that the field theory argument is not completely rigorous because the theory in question is not actually ever weakly coupled. On the contrary using the gravity dual description seems like a more reliable  approach here.
On the gravity side the fundamental matter is represented through a stack of D7-branes \cite{KarchKatz}.
In a general case when the number of flavors is comparable to the number of colors the analysis could be very complicated. Indeed to prove the existence of supersymmetric vacuum with, say, a given value of baryon vevs, one would have to construct a novel supergravity solution which includes backreaction of the D7-branes.
This is a complicated task and so far this was done only in a very limited case of the KS solution (i.e. for a special vacuum $\A=\B$) when D7-branes are smeared \cite{Benini}. At the same time whenever quenched approximation is valid the task simplifies significantly reducing it to the problem of constructing a solution for a supersymmetric D7 probe. Even this task could be complicated enough. Thus it is widely believed that the field theory discussed in \cite{Ouyang} does not experience SUSY breaking. Nevertheless the corresponding supersymmetric solution for the probe D7-brane
was proven difficult to find even in the simplest case of the $\A=\B$ KS vacuum  \cite{COS}. In this paper we focus on a model suggested by S.~Kuperstein in \cite{Kuperstein} who has shown that the corresponding field theory has supersymmetric vacuum with $\A=\B$ by embedding a supersymmetric D7-probe into the KS solution. In this paper we extend this analysis to the baryonic branch and construct the supersymmetric solution for D7-brane in probe approximation in the presence of a nontrivial baryionc condensate $\A\neq\B$. In this way we show that the baryonic branch of moduli space of the original theory survives after flavor sector is added in accordance with the field theory expectations.

To find supersymmetric probe configuration one could satisfy the first order kappa-symmetry condition rather than solving the second-order equations of motion. Despite the fact that the $\N=1$ background could be quite complicated
the kappa-symmetry condition always admits an elegant interpretation in term of generalized calibrations \cite{Martucci,MartucciII}. This formalism follows from the first order conditions that assure the supersymmetry of the background.
Employing the calibration formalism in this paper yields an interesting byproduct. Although the BGMPZ solutions were found by satisfying some of the first order supersymmetry constraints, not all of them were necessary to find the solutions. Considering the other ones  reveals some interesting and previously unknown properties of the BGMPZ solutions. For example the kappa-symmetry constraint for the D5-brane wrapping 3-cycle on the conifold helps to find the expression for the dilaton field through other
parameters of the solution (\ref{dilaton})  \cite{DKT}. In this paper we remind the reader the derivation of this result and also find some other interesting relations which can be useful for future studies of the BGMPZ backgrounds.

In a special case of the ISD background, like the KS solution, the kappa-symmetry constraint for the D7-brane admits an interesting generalization providing a new method to construct non-supersymmetric solutions for the
D7-probe \cite{DymarskyKuperstein}. In this paper we investigate the question whether this approach can be generalized for the non-ISD supersymmetric backgrounds by considering the example of the BGMPZ solutions.

This paper is organized as follows. We present a concise review of the KS and BGMPZ solutions in the next section followed by a discussion of the kappa-symmetry condition in section 3. We solve the kappa-symmetry equation for the
D7-brane and hence find the solution for the supersymmetric probe in section 4. Section 5 concludes with discussion.

\section{Review of the KS and BGMPZ backgrounds}

The solutions of the BGMPZ family can be thought of as a generalization of the KS solution. They all (including the KS solution which is a particular representative of the family) share the same complex structure but have different metric. Therefore we can use the same complex coordinates to describe  all solutions from the family.  In general the geometry is a warped product of the four dimensional Minkowski space and a six dimensional deformed conifold
\bea
\label{defcondeqn}
\sum_{i=1}^4 z_i^2=\epsilon^2\ ,
\eea
a smooth cone over $S^2\times S^3$.

In this paper we follow the notations of \cite{BGMPZ} who employ the PT ansatz \cite{PT} in the string frame.
Thus in what follows the deformation parameter $\epsilon$ is taken to be constant as defined in (\ref{norm}) unless we restore it explicitly to cancel the dimension of $\mu$.
With the exception of the KS solution which is a warped CY, the metric of the BGMPZ solutions is not Ricci flat but pseudo-Kahler
\bea
ds^2=e^{2A}dx^2+\sum_{i=1}^6 G_i^2\ .
\eea
One can define the one-forms $G_i$ in terms of the differential of the radial coordinate $dt$ and the one-forms $e_1,e_2$ and $\epsilon_1,\epsilon_2,g_5$ which form a basis on $S^2$ and $S^3$
\begin{eqnarray} \label{Gforms}
G_1\equiv e^{(x+g)/2}\,e_1\ ,&& \qquad
G_2\equiv {\cosh(t)+a\over \sinh(t)}\,e^{(x+g)/2}\,e_2 + {e^g \over \sinh(t)}\,e^{(x-g)/2}\,(\epsilon_2-a e_2)\ ,\nonumber\\
G_3\equiv e^{(x-g)/2}\,(\epsilon_1-a e_1)\ ,&&\qquad
G_4\equiv {e^g \over \sinh(t)}\,e^{(x+g)/2}\,e_2 - {\cosh(t)+a\over \sinh(t)}\,e^{(x-g)/2}\,(\epsilon_2-a e_2)\ ,\nonumber\\
G_5\equiv e^{x/2}\,v^{-1/2}dt\ ,&& \qquad G_6\equiv
e^{x/2}\,v^{-1/2}g_5\ .
\end{eqnarray}
The functions $a,g,x,v$ which are used in (\ref{Gforms}) to define metric, the warp factor $A$ and the dilaton $\phi$ depend only on radial coordinate $t$. We give the explicit relations which define these functions (as well as the functions $h_2,\chi,b$ which are used in the definition of fluxes) in the Appendix A.

The forms $\mathbb{G}_I=(G_{2I-1}+iG_{2I})$ are holomorphic and we can define a non-vanishing
$(3,0)$-form
\bea
\label{Omega}
\Omega = (G_1 + i G_2)\we (G_3 + i G_4)\we (G_5 + i G_6)=\\ \nonumber
-{ie^{3x/2}v^{-1/2}\over \sinh t}
(dt+ig_5)\wedge\left[ (e_1\wedge e_2+\epsilon_1 \wedge
\epsilon_2)\right. +\\ \nonumber
 \left. +i\sinh(t)(e_1\wedge
\epsilon_1+e_2\wedge \epsilon_2) +{\cosh(t)}(e_1\wedge
\epsilon_2+\epsilon_1\wedge e_2) \right]\ ,
\eea
and the warped fundamental $(1,1)$ form
\begin{equation}
J ={i\over 2} \Big [ (G_1 + i G_2) \we (G_1- i G_2) + (G_3 + i
G_4) \we (G_3- i G_4) + (G_5 + i G_6) \we (G_5- i G_6) \Big ]\ .
\label{J}
\end{equation}
As was mentioned above $J$ is not closed even if warping is removed. Since BGMPZ solutions share the same complex structure the $(3,0)$ form
 $\Omega$ should be proportional to a unique closed $(3,0)$ form of the Calabi-Yau (\ref{defcondeqn}).

There are also flux forms
\begin{eqnarray}
\nonumber
H &=& dB\ ,\qquad
B =B^0+\chi dg_5\ ,\qquad dg_5=(e_1\wedge e_2-\epsilon_1\wedge \epsilon_2)\ ,\\ \nonumber
B^0&=& h_2\left(\cosh t (\epsilon_1 \we \epsilon_2 + e_1 \we e_2)+(\epsilon_1 \we e_2 -  \epsilon_2 \we e_1 )\right)\ , \\ \nonumber
F_3 &=& -{1\over 2} g_5\we \big[  \epsilon_1 \we \epsilon_2  + e_1
\we e_2 - b\,  (\epsilon_1 \we e_2 - \epsilon_2 \we e_1) \big]
-{1\over 2}  dt \we \big[ b'\, (\epsilon_1 \we e_1 +
\epsilon_2 \we e_2) \big]\ ,\\ \nonumber
{F}_5 &=& {\cal F}_5  +  *_{10}{\cal F}_5\  , \qquad *_{10}{\cal F}_5=dx^0\we..\we dx^3\we d(e^{4A})\ ,\\
\label{equ:PTforms}
{\cal F}_5 &=& -h_2 (\cosh t + b)\, e_1 \we e_2 \we \epsilon_1 \we \epsilon_2
\we g_5\ .
\end{eqnarray}
Except for dilaton all other fields are zero.

The functions $a,g,x,v,\phi,A,h_2,\chi$ are not known explicitly. They can be expressed through $a,v$ and some functions of $t$ while the functions $a,v$ satisfy a system of first order differential equations. The relations presented in the Appendix A are sufficient to perform the calculations done in the next section.

The BGMPZ solutions are dual to the $SU((k+1)M)\times SU(kM)$ theory on the baryonic branch of moduli space i.e. they correspond to the different IR states of the same field theory. Correspondingly the BGMPZ solutions share the same behavior in the UV region where $t$ is large but differ in the IR $t\rightarrow 0$.

The first order system (\ref{av}) which defines the functions $a,v$ has one dimensional family of regular solutions. It can be parametrized by the subleading behavior at infinity
\bea
\label{aUV}
a\rightarrow -2e^{-t}+U(t-1)e^{-5t/3}+O(e^{-7t/3})\ .
\eea
Here real parameter $U$ is related to the expectation value of the baryon operators $\A,\B$ \cite{DKS,BDK}.

\section{Kappa symmetry}
\label{se:kappasymm}
The embedding of the D7-brane is specified by a four-cycle $\Sigma$  on the conifold.
The supersymmetry requires $\Sigma$ to be holomorphic and the kappa-symmetric equation to be satisfied by a world-volume gauge field. In this paper we focus on a particular embedding
\bea
\label{Kuembb}
z_4=\mu\ .
\eea
It was shown in \cite{Kuperstein} that the D7-brane embedded along (\ref{Kuembb}) satisfies the kappa-symmetric equation
with zero world-volume gauge field in the KS case. In this section we derive the explicit form of the kappa-symmetry equation in the case of a general background from the BGMPZ family.
In what follows we employ the formalism developed in \cite{Martucci} and also follow the notations introduced there.

The rigid structure of supersymmetry requires any $\N=1$ background to satisfy certain conditions
on the polyforms $\Psi_1,\Psi_2$ that are defined through the Killing spinor and encode information about geometry of the background.
In the case of baryonic branch backgrounds these conditions can be written as (see Appendix B for the derivation of $\tilde F$)
\bea
\label{cond1}
e^{-2A+\phi}(d+H\wedge)[e^{2A-\phi}\Psi_1]&=&dA\wedge \bar \Psi_1 +i {e^{A+\phi}}\tilde F\ ,\\
\label{cond2}
e^{-2A+\phi}(d+H\wedge)[e^{2A-\phi}\Psi_2]&=&0\ ,\\
\Psi_1={e^A(-\sqrt{1+e^{2\phi}}+ie^\phi)}e^{-i J}\ ,\quad
\Psi_2&=&-{ie^A}\Omega\ ,\nonumber \\ \qquad
\tilde F&=&4dA+e^{-2\phi}H\ . \nonumber
\eea
We start with the simpler equation (\ref{cond2}). It establishes that the $(3,0)$ form $e^{3A-\phi}\Omega$ is closed.
There is only one closed $(3,0)$ form on the conifold and therefore $e^{3A-\phi}\Omega$ should be proportional to it for each $U$. Since $e^{3A-\phi}\Omega$ has the same UV asymptotic for all $U$ we conclude that $e^{3A-\phi}\Omega$ is
actually  the same for all $U$ and from here follows the expression for the dilaton (\ref{dilaton}).
It can be shown that the probe action for the domain wall D5-brane wrapping a 3-cycle on the conifold is calibrated by $e^{3A-\phi}\Omega$ \cite{Martucci} and the calibration condition is saturated for all $U$ only when D5 is covering the minimal $S^3$ at the tip \cite{DKT}. Since the calibration form is the same for all $U$ the tension of the BPS domain wall is the same for all $U$ as well in accordance with the field theory expectations \cite{DKS}.

The equation (\ref{cond1}) implies the following identities which, as well as (\ref{dilaton}), can be checked straightforwardly\footnote{The similar relations were found for the backgrounds related to the BGMPZ family via a duality transformation in \cite{MM}.}
\bea
\label{constr1}
d(e^{2A}J)=UH\ ,\\
\label{constr2}
e^{2A-2\phi}H\wedge J=2 U dA\wedge J \wedge J\ .
\eea

As was shown in \cite{Martucci} the kappa-symmetry condition can be interpreted in terms of the generalized calibrations.
Indeed for the D7-brane the action is calibrated by some form $w$
\bea
\label{ineq}
S_{D7}=\int_\Sigma e^{4A-\phi}\sqrt{\text{det}(g+\F)}+e^{4A}\tilde C\wedge e^{\F}\geq\int_\Sigma w\ ,\\
\nonumber
w=\left. -e^{4A}\left[\Re\left(-ie^{-A-\phi}\Psi_1\right)-\tilde C\right]\wedge e^{\F}\right|_{top\ form\ on\ \Sigma}=\\
\label{wcal}
={e^{2A}J\wedge e^{2A}J\over 2}-U e^{2A}J\wedge \F+U^2B\wedge \F-{U^2 B\wedge B\over 2}\ .
\eea
Here $\F$ is the gauge-invariant flux on D7 $\F=B+d\gaugeA$.

Although (\ref{wcal}) defines $w$ only on $\Sigma$ is can be extrapolated outside of $\Sigma$ by continuing the world-volume gauge field $A$  outside of $\Sigma$ in an arbitrary way. Using (\ref{constr1},\ref{constr2}) one can check that (\ref{wcal}) defines a closed form. Then kappa-symmetry equation is the condition that the inequality (\ref{ineq}) is saturated
\bea
\nonumber
\left. \Im\left(ie^{A-\phi}\Psi_1\right)\wedge e^{\F}\right|_{top\ form\ on\ \Sigma}=\left.
{U \over 2}\left(J\wedge J-\F\wedge \F\right)+e^{2A}J\wedge \F\right|_{\Sigma}=0\ .\\
\label{ksc}
\eea
The expression above defines a 4-form on $\Sigma$ which must vanish for supersymmetry to be preserved.
As it is a form of highest degree on $\Sigma$ it is closed. Nevertheless even if continued outside of $\Sigma$ the 4-form (\ref{ksc}) is closed \cite{Martucci}. This in fact guarantees that the first-order kappa-symmetry equation can be reduced to an algebraic one similarly to the case of the Euclidean D5-brane on the conifold \cite{BDK}.

To find the supersymmetric solution for the probe D7-brane we have to find the world-volume gauge field $A$  satisfing  (\ref{ksc}). This is done in the next section. In the rest of this section we re-derive the kappa-symmetry equation (\ref{ksc}) and investigate whether it can be used to construct the non-SUSY solutions.

\subsection{Kappa-symmetry condition and minimum of action}
The kappa-symmetry equation can be thought of as a condition that minimizes the probe action.
This definition does not require supersymmetry to be explicitly involved and we use it here to demonstrate how one can arrive at
(\ref{ksc}) without knowing anything about kappa-symmetry transformation of world-volume fermions.
We do not have a goal to give a general derivation but rather to demonstrate the idea. Therefore we restrict our consideration to the case of the baryonic branch backgrounds.
We start with the probe action
\bea
\label{action1}\nonumber
S_{D7}=\int_\Sigma e^{4A-\phi}\sqrt{\text{det}(g+\F)}+e^{4A}\tilde C\wedge e^{\F}=\\ \int_\Sigma
e^{4A}\left[e^{-\phi}\sqrt{\text{det}(g+\F)}+{\F\wedge\F\over 2}+e^{-4A}U^2B\wedge\F-{e^{-4A}U^2\over 2}B\wedge B\right]\ ,
\eea
and notice that because of (\ref{constr1}) one can get rid of $B$ in (\ref{action1})
\bea
\label{action2}
S_{D7}=\int_\Sigma e^{4A}\left[e^{-\phi}\sqrt{\text{det}(g+\F)}-\left({J\wedge J\over 2}-{\F\wedge\F\over 2}\right)+e^{-2A}U J\wedge\F \right]+\int_\Sigma d(\dots)\ .
\eea
Although the world-volume $\Sigma$ of the D7-brane is not necessarily compact we neglect the integral of full derivative in (\ref{action2}) because it contributes only at infinity and hence can not change the local condition for
the embedding and world-volume gauge field. The expression above is algebraic in $J$ and $\F$. Therefore to minimize the integral we can try to minimize the expression inside the brackets at each point on $\Sigma$.
Let us choose the local coordinates $\sigma$ such that the induced metric $g$ at the given point becomes an identity matrix. Then the $SO(4)$ symmetry which respects the local form of the metric can be used to bring $J$ to the following form
\bea
J=\left(
\begin{array}{cccc}
0 & \cos\varphi & 0 & 0 \\
-\cos\varphi & 0 & 0 & 0 \\
0 & 0 & 0 & 1 \\
0 & 0 & -1 & 0
\end{array}\right)\ .
\eea
This is in agreement with the fact that the volume form on $\Sigma$, which is in our coordinate basis is just $d^4\sigma$, is always bigger or equal than ${J\wedge J\over 2}=d^4\sigma\cos^2{\varphi\over 2}$.
The antisymmetric matrix $\F$ can be represented as the two vectors in the three-dimensional space -- the representation ${\bf 3}\oplus {\bf 3}$ of the symmetry group $so(3)\times so(3)\cong so(4)$. For the given lengths $r_a,r_s$ these vectors must be aligned along the same direction with $J$ to minimize (\ref{action2}). Therefore the field $\F$ is
\bea
\F=-\left(
\begin{array}{cccc}
0 & r_s+r_a & 0 & 0 \\
-r_s-r_a & 0 & 0 & 0 \\
0 & 0 & 0 & r_s-r_a \\
0 & 0 & r_a-r_s & 0
\end{array}\right) \ .
\eea
The expression in the brackets from (\ref{action2}) becomes
\bea
\label{bracket}
L=\cosh \psi \sqrt{1+2(r_a^2+r_s^2)+(r_a^2-r_s^2)^2}-(\cos\varphi+r_a^2-r_s^2)-\\
\nonumber -\sinh\psi(r_s(1+\cos\varphi)+r_a(1-\cos\varphi))\ ,\\
\cosh\psi=e^{-\phi}\geq 1\ .
\eea\
This expression is non negative. Indeed at the minimum $\cos \varphi$ is either $1$ or $-1$ depending on the sign of
$1+\sinh\psi(r_s-r_a)$. In the first case when $1+\sinh\psi(r_s-r_a)>0$ and $\cos\varphi=1$ the expression for $L$ becomes
\bea
\label{isdL}
L=\cosh\psi \sqrt{x^2+y^2}-x-\sinh\psi\ y\ ,\\
x=1+r_a^2-r_s^2\ ,\quad y=2r_s\ .
\eea
This expression reaches its minimum $L=0$ when $y=\sinh\psi\ x$ (this condition is compatible with $1+\sinh\psi(r_s-r_a)>0$).
The constraint $\cos\varphi=1$ is equivalent to the condition that $\Sigma$ is holomorphic and $y=\sinh\psi\ x$ is exactly the kappa-symmetry equation (\ref{ksc}).

Whenever $1+\sinh\psi(r_s-r_a)<0$  and $\cos\varphi$ approaches $-1$ one can change the orientation on $\Sigma$ effectively changing sign of $\cos\varphi$ and interchanging $r_s$ and $r_a$. This would reduce the problem to the previously considered case (\ref{isdL}) with $x\rightarrow -x$. Thus we have shown that the kappa-symmetry condition that the embedding is holomorphic and the world-volume field satisfies the equation (\ref{ksc}) can be recovered by minimizing the probe action.

One can invert the order and minimize the action (\ref{action2}) with respect to the gauge field first and then minimize the result with respect to the embedding $\varphi$. This is particularly easy to do in the special case of the ISD solution when dilaton vanishes $\psi=0$
\bea
L=\sqrt{1+2(r_a^2+r_s^2)+(r_a^2-r_s^2)^2}-(\cos\varphi+r_a^2-r_s^2)\ .
\eea
In this case the constraint that $\F$ is anti-self-dual (i.e. $r_s=0$) minimizes $L$ with respect to $r_s,r_a$ for any embedding $\varphi$. This observation was used in \cite{DymarskyKuperstein} to propose a  method to construct the non-SUSY solutions: one is to find the embedding $\Sigma$ which extremizes  the effective action $L_{eff}=1-\cos\varphi$
which is simply (second term can be dropped because $e^{2A}J$ is closed in this case)
\bea
\int_\Sigma \sqrt{g}-\int_\Sigma e^{4A}{J\wedge J\over 2}\ ,
\eea
and then to find an anti-self-dual $\F$ that satisfies the Bianchi identity. This method was used in \cite{DymarskyKuperstein} to find a new non-supersymmetric solution for the probe
D7-brane embedded in the KS background.

Naturally we would like  to generalize this approach to a general background with $\psi\neq 0$.
The idea would be the same -- to minimize the probe action (\ref{action2}) with respect to the gauge field for a given (not necessarily holomorphic) embedding. As a result one will obtain an effective action that depends only on the embedding.
If this action admits a non-holomorphic surface as a solution one can build a non-supersymmetric solution for the D-brane by accompanying the embedding by an appropriate gauge field. An obvious advantage of this approach  compared with the conventional one of solving EOM is that the equation for the gauge field resulting from minimization of the action (\ref{action2}) is a first order one.

Now we will investigate the possibility to apply this method in the case of baryonic branch of the KS solution.
We start with the KS background ($\psi=0$) and move along the baryonic branch a little bit such that $\psi$ is very small but non-zero. Then we try to find the minimum of (\ref{bracket}) expanded to the first non-trivial order in $\psi$
\bea
\label{linearL}
L=\sqrt{1+2(r_a^2+r_s^2)+(r_a^2-r_s^2)^2}-(\cos\varphi+r_a^2-r_s^2)-\\ -\psi(r_s(1+\cos\varphi)+r_a(1-\cos\varphi))+O(\psi^2)\ ,\nonumber
\eea
with respect to $r_a,r_s$ for the given $\varphi$. If $\cos\varphi \neq 1$ effective action (\ref{linearL}) is minimized when $r_s\rightarrow 0$ and $r_a\rightarrow \infty$. In fact the action does not approach $-\infty$ as it may seem from (\ref{linearL}).  As we know the action is bounded by zero by below. The  approximate expression (\ref{linearL}) is simply not applicable anymore when $r_a$ is bigger than $1/\psi$. But it is important that $r_a$ of order $1$ is not a minimum. Therefore, unless $\cos\varphi=1$, the expression for $r_a,r_s$ which minimizes the action (\ref{bracket}) is not smooth as a function of $\psi$ near $\psi=0$. This suggests that the condition on $\F$ to minimize the action for the given embedding may not be well-defined.

On the contrary if $\cos\varphi=1$ the minimum can be found with help of the ``kappa-symmetry equation''  $y=\sinh \psi\ x$. This equation admits continuous solution $s\sim O(\psi)$ and $r_a\sim  1+O(\psi)$ for small $\psi$ near $\psi=0$ and can be used to find the (necessarily supersymmetric) solutions for D-brane probe.

The analysis for small $\psi$ above can be generalized for any $\psi$. We rewrite (\ref{bracket}) as follows
\bea
\label{xyz}
L=\cosh\psi \sqrt{x^2+y^2+z^2}-x-\sinh\psi\ y\ ,\quad \quad \quad \quad \quad \quad \\ \nonumber
x=\cos\varphi +r_a^2-r_s^2\ ,\quad y=(1+\cos\varphi)r_s+(1-\cos\varphi)r_a\ ,\quad z=\sin\varphi\sqrt{1+(r_a-r_s)^2}\ .
\eea
Clearly this expression is non-negative and it reaches its minimum $L=0$ when
\bea
y=\sinh\psi x\ ,\quad z=0\ .
\eea
Unless $\varphi=0$ or $r_a=r_s$ the minimum is reached at infinity when $x$ and $y$ satisfy
$y=\sinh\psi x$ and much larger than $z$. Thus we confirmed our initial impression that the condition
on gauge field $\F$ that minimizes the action (\ref{bracket}) does not lead to a well-defined $\F$ unless the embedding is holomorphic $\varphi=0$ or the background is ISD when dilaton is zero. Another option is to require $r_a=r_s$ and then
the action is minimized when $r_a=r_s=\sinh\psi \cos\varphi/2$. This solution though looks too restrictive and it could be impossible to satisfy the Bianchi identity $d\F=H$ with such $\F$.

Our failure to find a condition for $\F$ to minimize (\ref{bracket}) and in this way develop a method to construct non-SUSY solutions for D-brane probe does not necessarily mean this is impossible. Our idea above was to minimize the action with respect to $\F$, In fact to find any condition which solves $\partial L /\partial \F=0$ for general embedding would be enough. We leave the task to investigate this question for the future but would like to note here that even if there is no any trick available to find the non-supersymmetric solutions via solving algebraic equation for $\F$ it does not mean the solution
of \cite{DymarskyKuperstein} ceases to exist on the baryonic branch away from the KS point. It  would rather mean
that to find these non-supersymmetric solutions one would have to solve the second-order supergravity EOM.

\section{Solution for fluxes}
\label{se:solution}
In this section we proceed with the kappa-symmetry equation for the world-volume gauge field (\ref{ksc}) and solve it
for the case of the D7-brane embedded along the Kuperstein embedding (\ref{Kuembb}).
So far in section (\ref{se:kappasymm}) we used the general properties of the solution dictated by supersymmetry, like (\ref{constr1},\ref{constr2}). To solve the kappa-symmetry solution
we would need to use particular properties of the BGMPZ solutions described below.
Thus the crucial observation is that the  relation (\ref{constr1}) can be integrated yielding a simple form
\bea
e^{2A}J=UB-d[(\lambda +U\chi)g_5]\ ,\\ \nonumber
\lambda=U{e^{2\phi}a(t\cosh t-\sinh t)\over 2 (a\cosh t+1)}\ .
\eea
The function $\lambda$ has a nice property
\bea
{d\over dt}\left(\lambda+U\chi\right)=v^{-1}e^{2A+x}\ ,
\eea
and therefore
\bea
\label{Jexp}
e^{2A}J=U B^0+v^{-1}e^{2A+x} dt\wedge g_5-\lambda dg_5\ .
\eea
Such a  simple form of $J$ suggests the anastz for the world-volume gauge field
\bea
\label{ansatz}
\gaugeA=\xi(t) g_5\ ,\\
\label{Fexp}
\F=B^0+(\xi+\chi)dg_5+\xi' dt\wedge g_5\ .
\eea
Let us note here that the ansatz (\ref{ansatz}) preserves the full $SU(2)\times SU(2)$ symmetry of conifold while the
D7 embedding (\ref{Kuembb}) breaks it to the diagonal $SO(3)$. Therefore our ansatz looks too restrictive at this moment. Nevertheless we will be able to solve the kappa-symmetry equation even with the simplest ansatz (\ref{ansatz}).

After plugging (\ref{Jexp},\ref{Fexp}) into (\ref{ksc}) and using that (see Appendix C for the proof)
\bea
\left.dt\wedge g_5\wedge B^0\right|_{\Sigma}=0\ ,
\eea
we arrive at the form equation on $\Sigma$
\bea
\nonumber
dt\wedge g_5\wedge dg_5 \left[\xi'(\xi+\chi)-{e^x\over v\sqrt{e^{-2\phi}-1}}(\xi+\lambda)+{\xi' \lambda\over U}+{\sqrt{e^{-2\phi}-1}e^x \lambda\over v U }\right]+\\
\label{kappas}
e_1\wedge e_2\wedge \epsilon_1\wedge \epsilon_2\left[{\lambda^2(e^{-2\phi}-1)\over U^2}-{2\lambda\over U}(\xi+\chi)-(\xi+\chi)^2-e^{-2\phi}h_2^2\sinh^2 t\right]=0\ .
\eea
Both forms $dt\wedge g_5\wedge dg_5$ and $e_1\wedge e_2\wedge \epsilon_1\wedge \epsilon_2$ are the $SO(3)$
invariant forms of highest dimension on $\Sigma$. Therefore they are proportional to each other with the coefficient which may depend only on radius $t$ but not on the angles
\bea
\label{fdef}
\left. e_1\wedge e_2\wedge \epsilon_1\wedge \epsilon_2\right|_{\Sigma}=f(t)\left. dt\wedge g_5\wedge dg_5 \right|_{\Sigma}\ .
\eea
Let us define $\a(t)$ through
\bea
\label{af}
{\a'\over \a}=-2f\ .
\eea
Then the equation (\ref{kappas}) can be integrated as follows
\bea
\label{xiequation}
-{1\over \a}{d\over dt}\left[\a\left((\xi+\chi)^2+{2\lambda \over U}(\xi+\chi)+\left(e^{-2\phi}h_2^2\sinh^2 t-{\lambda^2\over U^2}(e^{-2\phi}-1)\right)\right)\right]=0\ .
\eea
The fact that the kappa-symmetry equation can be integrated to the algebraic equation is not surprising
as was discussed in section \ref{se:kappasymm}.
Similarly the on-shell action density should be a full derivative as follows from (\ref{wcal}).
We plug (\ref{Jexp},\ref{Fexp}) into (\ref{wcal}) to get
\bea
\label{effaction}
S_{D7}=\int_\Sigma dt\wedge g_5\wedge dg_5\ {1\over \a} {d\over dt}\left[\a\left(U\xi(\lambda+U\chi)+{(\lambda+U\chi)^2\over 2}\right)\right]\ .
\eea
The only way for (\ref{effaction}) to be an integral of full derivative is to equate $a(t)$ to the volume of the D7-profile $S^3\subset \Sigma$ at fixed radius $t$
\bea
\label{a}
\int\limits_{S^3=\Sigma\ at\ fixed\ t}g_5\wedge dg_5=\a(t)=32\pi^2 {\epsilon^2\sinh^2 t-2|\mu^2|\cosh t +\mu^2+\bar \mu^2\over \epsilon ^2 \sinh^2 t}\ .
\eea
Indeed one can check that this definition of $\a$ agrees with (\ref{af}) as shown in the Appendix C.

The function $\a(t)$ vanishes at the tip of D7-brane where $|\epsilon^2-\mu^2|=\epsilon^2\cosh t-|\mu^2|$.
Therefore (\ref{xiequation}) should be integrated with zero integration constant
\bea
(\xi+\chi)^2+{2\lambda\over U}(\xi+\chi)+\left(e^{-2\phi}h_2^2\sinh^2 t-{\lambda^2\over U^2}(e^{-2\phi}-1)\right)=0\ .
\eea
Only one root  of this quadratic equation is regular at $t\rightarrow \infty$ and $U\rightarrow 0$.
Indeed when $t\rightarrow \infty$
\bea
\lambda \rightarrow -e^{2t/3}+{U\over 2}(t-1)+O(e^{-2t/3})\ ,
\eea
while
\bea
\left(e^{-2\phi}h_2^2\sinh^2 t-{\lambda^2\over U^2}(e^{-2\phi}-1)\right) \rightarrow O(t^2)\ .
\eea
Therefore the physical $\xi$ is given by
\bea
\xi+\chi={-\lambda+e^{-\phi}\sqrt{\lambda^2-U^2 h_2^2\sinh^2 t}\over U} \rightarrow UO(t^2)e^{-2t/3}\ .
\eea
The bulk field $\chi$ has a similar asymptotic behavior in the UV: $\chi \rightarrow UO(t)e^{-2t/3}$.

Now we can estimate the on-shell value of the action (\ref{effaction}) calculated at some cut-off scale $t^*$.
In the far UV where we can neglect the subleading $O(e^{-2t^*/3})$ terms it approaches
\bea
\label{onshellaction}
S_{D7}=32\pi^2 \left(e^{4t^*/3}+U^2O(t^{*2})\right)\ .
\eea
The leading term $e^{4t^*/3}$ term is $U$-independent. It corresponds to the geometrical volume of the D7-brane.
The subleading $U$-dependent term is logarithmically divergent (if expressed through the conventional
radius variable $r^3=\epsilon^2 \cosh t$). Both divergent terms should be canceled  by a proper renormalization  procedure as required by unbroken supersymmetry. It is interesting to note that  (\ref{onshellaction}) has a simple $U$ dependence similar to those of D3-brane placed on the baryonic branch \cite{DKS}. When the throat is compactified we expect that the $U$-dependent term from (\ref{onshellaction}) will generate a potential along the branch proportional to $U^2$.

\section{Discussion}
In this paper we found the world-volume gauge field on the D7-brane
embedded along the Kuperstein embedding  $z_4=\mu$ in a  baryonic branch background of \cite{BGMPZ} which satisfies the kappa-symmetry equation. Hence we demonstrated that the whole baryonic branch of moduli space survives after the fundamental matter is added to the original $\N=1$ $SU((k+1)M)\times SU(kM)$ theory.

The fact that the D7-brane can be placed on the baryonic branch without breaking supersymmetry is interesting in the context of various phenomenological models based on the compactification of the KS solution. Thus a wide range of stringy models of inflation employ the KS solution as an appropriate approximation to the throat region of a compact manifold. In these scenarios the D7-brane plays a crucial role of stabilizing the size of the compactification manifold through the non-perturbative effect of gaugino condensation \cite{KKLT}. To assure stability of the D7-brane (and flat potential for the inflaton field) it should be supersymmetric (in the non-compact limit). Now this construction can be generalized for the backgrounds from the BGMPZ family. One advantage of this construction is a mechanism to tune the potential of the probe D3-brane in the scenario of \cite{KKLMMT} by choosing an appropriate value of the baryonic condensate \cite{DKS}.  Note though that this in turn would require a fined tuned potential for $U$ which could be difficult to achieve. Another interesting scenario is to consider a closed string inflation of the baryonic vev $U$ after the baryonic branch is uplifted by the compactification effects.

\vskip1cm

I am grateful to S.~Kachru, I.~R.~Klebanov, J.~Maldacena, L.~Martucci, and Y.~Tachikawa for useful discussions.
This work is supported by Stanford Institute for Theoretical Physics and in part by the grants NSh-3035.2008.2 and RFBR 07-02-00878.

\appendix
\section{BGMPZ solutions}
The functions $a,v$ are defined through the system of first order differential equations
\bea
\label{av}
a'&=&- {{\sqrt{-1 - a^2 - 2 a \cosh t}} \left( 1 + a\cosh t
\right)\over v  \sinh t}   -
  {a\sinh t \left( t + a \sinh t \right) \over t \cosh t - \sinh t}\ , \\
  \nonumber
v' &=& \displaystyle {-3\,a\,\sinh t\over{\sqrt{-1 - a^2 -2 a \cosh t}}} +\\
      &+& v \, \left[- a^2 \cosh^3 t + 2\,a\,t \coth t +
       a \cosh^2 t \left( 2 - 4\,t \coth t \right) +
       \cosh t \left( 1 + 2\,a^2 \right. \right.  \nonumber \\
      &-&\left. \left. \left( 2 + a^2 \right) t \coth t \right)
           +  \, {t\over \sinh t} \right]/ \left[ \left( 1 + a^2 + 2 a \cosh t
       \right) \left( t \cosh t - \sinh t \right) \right]\ . \nonumber
\eea
There is one dimensional family of regular solutions parametrized by $U$ via the subleading behavior at infinity (\ref{aUV}). All other functions (except for $\chi$ which is defined through its $t$ derivative) can be expressed through $a,v$ and radius $t$ \cite{BGMPZ}
\bea
\nonumber
e^{2g}&=&-1-a^2+2ac\ ,\\
\nonumber
e^{2x}&=& {(bc-1)^2\over 4(ac-1)^2} e^{2g+2\phi}(1-e^{2\phi})\ , \\
\nonumber
b&=&-{t\over \sinh(t)}\ ,\\
\nonumber
h_2&=& {e^{2\phi}(bc-1)\over 2s}\ ,\\
\nonumber
\chi'&=& a(b-c)(ac-1)e^{2(\phi-g)}\ ,\\
c&=&-\cosh(t)\ ,\qquad  s=-\sinh(t)\ ,\qquad  g_s=1\ ,\qquad M\alpha'=2\ .
\eea
The warp function $A$ can be expressed through dilaton \cite{DKS}
\bea
\label{Aphi}
e^{-4A}=U^{-2}(e^{-2\phi}-1)\ .
\eea
Here we take $\epsilon$ to be a numerical constant such that the coefficient $\gamma$ defined in \cite{DKS} is
\bea
\label{norm}
\gamma=2^{10/3}(g_sM\alpha')^2\epsilon^{-8/3}=1\ .
\eea
The explicit expression for the dilaton \cite{DKT}
\bea
\label{dilaton}
e^{4\phi}=-{64v(a\cosh(t)+1)^3\sinh(t)^5\over
3U^3(-1-a^2-2a\cosh(t))^{3/2}(t\cosh(t)-\sinh(t))^3}\ ,
\eea
follows from the fact that $e^{3A-\phi}\Omega$ is $U$-independent as explained in section \ref{se:kappasymm} and therefore can be equated with the explicit Ks expression.

\section{Calculation of R-R fields}

In this section we use the notations of \cite{Martucci} and calculate the RR potentials that enter the Chern-Simons term
\bea
S_{CS}={\rm Vol_4}\int_{\Sigma} e^{4A}\tilde{C}\wedge e^{\F}\ ,\\
\nonumber
{\rm Vol_4}=\int {\rm vol_4}=\int d^4 x\ .
\eea
Here we divide the RR potentials into four and six-dimensional parts
\bea
C_k=\hat{C}_k+{\rm vol_4} e^{4A}\wedge \tilde{C}_{k-4}\ .
\eea
In what follows we
use the self-duality of field-strength
\bea
F_{k+1}=dC_k+H\wedge C_{k-2}=\hat F_{k+1} +{\rm vol_4}e^{4A}\wedge \tilde F_{k-3}\, \qquad
\tilde{F}_{k}=(-)^{(k-1)(k-2)\over 2}*_6\hat{F}_{6-k}\
\eea
to calculate $\tilde C$.

The form of the four-form RR-flux (\ref{equ:PTforms}) implies
\bea
\tilde C_0=1\ .
\eea
Using  the expression for the warp factor (\ref{Aphi})
and the relation between the R-R and NS-NS forms that holds for the whole baryonic branch\footnote{This
is not a ISD condition $*_6 F_3=-e^{\phi}H$. The latter is not satisfied for the BGMPZ family away from the KS solution due to running dilaton.}
\bea
*_6 F_3=-e^{-2\phi}H\ ,
\eea
we have
\bea
\tilde{F}_3=e^{-2\phi}H=e^{-4A}d(e^{4A}\tilde{C}_2)+H\ ,
\eea
and therefore 
\bea
d(e^{4A}\tilde{C}_2)=U^2H\ ,\quad
\tilde{C}_2=U^2e^{-4A}B\ .
\eea
Since the axion vanishes $C_0=0$ we also have
\bea
e^{4A}\tilde{F}_5=d(e^{4A}\tilde{C}_4)+e^{4A}H\wedge \tilde{C}_2=0\ ,\quad
\tilde{C_4}=-{U^2e^{-4A}\over 2}B\wedge B\ .
\eea
Eventually we have for the polyforms $\tilde C$ and $\tilde F$
\bea
e^{4A}\tilde C=e^{4A}+U^2B-{U^2\over 2}B\wedge B\ ,\\
\tilde F=4 dA+e^{-2\phi}H\ .
\eea

\section{External forms on D7 world-volume}
In this appendix we will prove several useful relations for the forms living on $\Sigma$
\bea
\label{emb}
z_4=\mu\ .
\eea

The calculation in section \ref{se:solution}
was done in basis $e_1,..,\epsilon_2,g_5$ while the embedding equation (\ref{emb}) is conveniently formulated in terms of homogeneous coordinated on the conifold $z_i$.
It is actually quite difficult to express (\ref{emb}) in terms of $e_1,..,\epsilon_2,g_5$.
Therefore we formulate the relations for the forms on $\Sigma$ in terms of the one-forms
$e_1,..,\epsilon_2,g_5$ but use the homogeneous coordinates to prove them.

We start with the relation between $e_1,..,\epsilon_2,g_5$ and $dz_i$ \cite{Remarks}
\bea
\label{g5}
\bar z_\mu dz_\mu=\epsilon^2 {\sinh t\over 2}(dt+ig_5)\ ,\\
\label{B_0}
{\texttt{B}}^0=\left( (\epsilon_1 \we \epsilon_2 + e_1 \we e_2)+{1\over \cosh t }(\epsilon_1 \we e_2 -  \epsilon_2 \we e_1 )\right)= {2i \epsilon_{\mu_1,..,\mu_4} z_{\mu_1} \bar z_{\mu_2} dz_{\mu_3}\wedge d\bar z_{\mu_4}\over \sinh t\cosh t}\ .
\eea
The form $dg_5$ can be obtained from (\ref{g5}).
Now we calculate the pullback on $\Sigma $ by substituting   $dz_4=0$ and $dz_3=-(z_1 dz_1+z_2 dz_2)/ z_3$ into the expressions above. As a result we express $dt\wedge g_5, dg_5$ and ${\texttt{B}}^0$ though $dz_1,dz_2$ and their complex conjugate.
To make sure  that we are on the right track we can check the relation between ${\texttt{B}}^0$ and $dg_5$
\bea
\left.\left.{\cosh^2 t\over \sinh^2 t}{\texttt{B}}^0\wedge {\texttt{B}}^0\right|_{\Sigma}=-dg_5\wedge dg_5\right|_{\Sigma}=2\ {4 |\mu\cosh t-\bar \mu|^2\over \epsilon^4 \sinh^4 t |z_3|^2}dz_1\wedge d\bar z_1\wedge dz_2\wedge d\bar z_2\ .
\eea
At the next step we calculate the pullback of $dt\wedge g_5\wedge {\texttt{B}}^0$ and find that it vanishes
\bea
\left. dt\wedge g_5\wedge {\texttt{B}}^0\right|_{\Sigma}=0\ .
\eea

Eventually we calculate the pullback of $dt\wedge g_5\wedge dg_5$ and find $f(t)$ (\ref{fdef})
\bea
\nonumber
\left.\left. -{f^{-1}\over 2}dg_5\wedge dg_5 \right|_{\Sigma} =dt\wedge g_5\wedge dg_5\right|_{\Sigma}\ ,\\
\label{fexpr}
f(t)=-{|\mu\cosh t -\bar \mu|^2\over \sinh t (\epsilon^2\sinh^2 t-2|\mu^2|\cosh t +\mu^2+\bar \mu^2)} \ .
\eea
The expression for $\a$ (\ref{a}) follows from (\ref{fexpr}) after integration.

\end{document}